\documentclass[prl,twocolumn]{revtex4} % showpacs 
\usepackage{amssymb}
\usepackage{amsmath}
\usepackage[dvipdfm]{graphicx}
\usepackage[colorlinks]{hyperref} %links for citations

\newcommand{\ket}[1]{\left | #1 \right \rangle}
\newcommand{\bra}[1]{\left \langle #1 \right |}
\newcommand{\kb}[1]{| #1 \rangle \langle #1 |}

\newcommand{\beq}{\begin{equation}}
\newcommand{\eeq}{\end{equation}}
\newcommand{\beqa}{\begin{eqnarray}}
\newcommand{\eeqa}{\end{eqnarray}}

\begin{document}
\title{Quantum computation on the edge of a symmetry-protected topological order}
\author{Akimasa  Miyake }
\email{amiyake@perimeterinstitute.ca}
\affiliation{%
\mbox{Perimeter Institute for Theoretical Physics, 31 Caroline Street North,
Waterloo Ontario, N2L 2Y5, Canada} }
\date{March 24, 2010} % July 6, 2009 --- 

\begin{abstract}
We elaborate the idea of quantum computation through measuring the correlation of a
gapped ground state, while the bulk Hamiltonian is utilized to stabilize the resource.
A simple computational primitive, by pulling out a single spin adiabatically from the bulk
followed by its measurement, is shown to make any ground state of the one-dimensional
isotropic Haldane phase useful ubiquitously as a quantum logical wire.
The primitive is compatible with certain discrete symmetries that protect this topological order, 
and the antiferromagnetic Heisenberg spin-1 finite chain is practically available. 
Our approach manifests a holographic principle in that the 
logical information of a universal quantum computer can be written and processed perfectly 
on the edge state (i.e., boundary) of the system, supported by the persistent entanglement 
from the bulk even when the ground state and its evolution cannot be exactly analyzed.

\end{abstract}

\maketitle
%%%%%%%%%%%%%%%%%%%%%%%%%%%%%%%%%%%%%%%%%%%%%%%%%
{\it Introduction.---}
A quantum computer is ``high maintenance,'' according to common folklore. 
In any possible architecture, it is supposed to be inevitable to engineer and control 
the quantum system in a fine precision humanity has yet to master,
despite a lot of brilliant ideas to build it practically and robustly.
The current paper follows the direction of the measurement-based quantum computation (MQC),
whose basic strategy can be said to simulate the time evolution of the logical qubits 
(i.e., the quantum logical wires in the language of the quantum circuit model) by measuring 
entanglement locally and sequentially \cite{raussendorf01}.
In particular our work is motivated by a ground-code scheme of MQC \cite{brennen08}, 
that utilizes a gapped ground state of the so-called 1D Haldane phase \cite{haldane83} for the sake 
of both quantum computation and a passive Hamiltonian protection.
Here we model this quantum phase primarily in terms of the 1D antiferromagnetic spin-1 chain 
of the length $N$ described by the isotropic, nearest-neighboring two-body Hamiltonian,
\begin{equation}
H = J \sum_{k=1}^{N-1} [{\mathbf S}_{k} \cdot {\mathbf S}_{k+1} - \beta
({\mathbf S}_{k} \cdot {\mathbf S}_{k+1})^2], 
\label{eq:H}
\end{equation}
where $J > 0$, and ${\mathbf S}_{k}$ is the spin-$1$ irreducible representation of $\mathfrak{su}(2)$
at the site $k$.
It has been well established that the system is in the gapped Haldane phase in the
region of $ -1 <\beta < 1$. The Heisenberg point at $\beta = 0$ is most significant in practice,
and the Affleck-Kennedy-Lieb-Tasaki (AKLT) point \cite{AKLT87} at $\beta = - \tfrac{1}{3}$ was utilized 
in \cite{brennen08}, since the AKLT ground state can be exactly 
described as a matrix product state (MPS) \cite{MPS92}.
In the general point in the Haldane phase, there has been known no exact compact
description of the ground state, although its numerical approximation by MPS, which is indeed the basis of
the density matrix renormalization group, is efficiently available
because of the gap independent of the system size $N$.
Regarding the boundary condition, an additional spin-$\tfrac{1}{2}$ degree ${\mathbf s}$ of 
freedom, coupled by the interaction $ \tfrac{4 J}{3} {\mathbf S} \cdot {\mathbf s}$ to each end 
($k= 1$ or $N$) of the bulk chain by spin $1$'s, was assumed in \cite{brennen08} for mathematical simplicity,
i.e., to ensure the uniqueness of the ground state for its preparation, while the two-fold degeneracy
during the computation.

\begin{figure}[b]
\includegraphics[width=6.0cm]{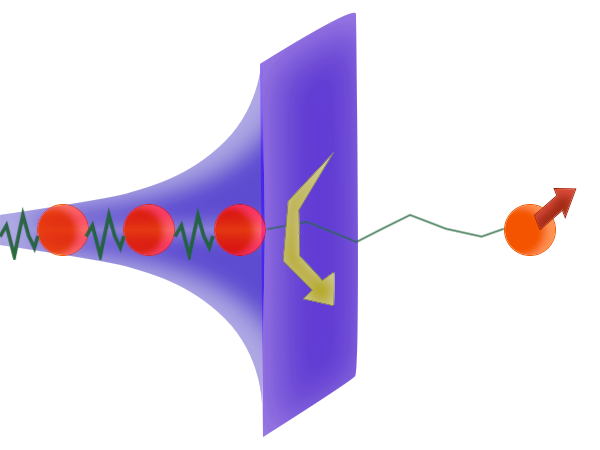} % 5.6 cm
\caption{Quantum computation is processed on the edge state which plays the role of a ``holographic screen,''
while its computational capability relies on the symmetry-protected topological entanglement from the bulk.}
\label{fig:hologram}
\end{figure}

Our motivation was, originally, to analyze to what extent the resource 
requirement can be relaxed. Intriguingly the analysis uncovers that
a simple operational primitive of the ground-code MQC can work {\it perfectly} regardless
of the choice of the ground state (i.e. the parameter $\beta$) in the Haldane phase
as well as without engineering the additional boundary terms with the artificial spin $\tfrac{1}{2}$'s. 
Thus, the Heisenberg finite chain with the open boundary is sufficient in practice.
In other words, we observe low-maintenance features of the ground-code MQC in that
this computation is doable without an exact (classical) description of the resource ground state 
as well as without an initialization to a pure state.
It turns out these features are deeply intertwined with the physics of the 1D Haldane phase 
(cf. Fig.~\ref{fig:hologram}), that is well characterized as the symmmetry-protected topological order 
in a modern perspective \cite{gu09,pollmann09}.
We believe our approach must bring the study of MQC, conventionally based on 
the analysis of the model entangled states (e.g., \cite{raussendorf01,verstraete04,gross07}), 
much closer to the condensed matter physics, which is aimed to describe characteristic physics based
on the Hamiltonian.

{\it Properties of the Haldane ground state.---}
In order to describe the general ground state for any $ \beta \in (-1, 1)$, we utilize a notion of the so-called edge 
state, a fractionalized spin-$\tfrac{1}{2}$ degree of freedom, {\it emergent collectively} in each end of 
the chain in the open boundary condition \cite{kennedy90}.
Its validity has also been confirmed experimentally \cite{hagiwara90}. 
Let us denote the quantum numbers of the total spin operators as
$(\sum_{k= j}^{N} {\mathbf S}_k)^2 = S_{\rm tot} (S_{\rm tot} + 1)$, 
$\sum_{k = j}^{N} {S}_k^{\alpha} = S_{\rm tot}^{\alpha}  \; (\alpha = x,y,z) $,
where the subscript ``tot(al)'' emphasizes they are the bulk quantities of
the remaining unmeasured chain once the computation gets into its $j$-th step. 
These two emergent spin $\tfrac{1}{2}$'s on the two ends of the chain explain heuristically that
the ground state in the periodic boundary is $S_{\rm tot} = 0$, by the pair  
making the global singlet, while the ground state in the open boundary 
consists of their global singlet and triplet of $S_{\rm tot} = 1$ \cite{kennedy90}, and thus
is four-fold degenerate.

As described in the next section, our computation is carried acting only on the one side of
the chain from the site $k=1$ toward $k=N$.
Note the edge state is the emergent degree of freedom localized near the boundary 
in a scale of the finite correlation length $\xi$ (see, e.g., \cite{polizzi98}).
So we can focus only on the edge state on the boundary where our measurements take place
and treat the ground state as if it has $S_{\rm tot} = \tfrac{1}{2}$ by the half-infinite chain,
in effectively ignoring the other edge state as far as $N \gg \xi$.
This argument is complemented by the later explanation about the open-chain scheme with explicit 
four-fold degeneracy, which indicates it is always possible to keep two edge states sufficiently distant 
if $N$ is initially large enough. 
Accordingly, in defining the logical computational basis spanned by
$\ket{{\mathcal G}_0} = \ket{S_{\rm tot} = \tfrac{1}{2}; S_{\rm tot}^z = \tfrac{1}{2}}$ and 
$\ket{{\mathcal G}_1} = \ket{S_{\rm tot} = \tfrac{1}{2}; S_{\rm tot}^z = -\tfrac{1}{2}}$,
the logical information stored in the ground subspace before the $j$-th measurement is described as 
$\ket{\Psi(j)} := a_0 \ket{{\mathcal G}_{0} (j)} + a_1 \ket{{\mathcal G}_{1} (j)} $.
Note that although the logical basis and operations formally vary by the computational step $j$, 
they should be interpreted as acting effectively on the edge state newly emergent in the remaining chain
of the length $N-j$.

{\it Computational primitive.---}
The primitive of the ground-code MQC is simple; at the step $j$, a certain measurement of
the single spin ${\mathbf S}_j$, following the adiabatic turning off of the two-body interaction 
acting on it (i.e., 
$h_{j, j+1} = J [{\mathbf S}_{j} \cdot {\mathbf S}_{j+1} - \beta ({\mathbf S}_{j} \cdot {\mathbf S}_{j+1})^2]$ 
of Eq.~(\ref{eq:H})).
The basis of the measurement is the control parameter, which is determined according to
the target algorithm, and the sequence of such primitives must be capable of
simulating any time evolution of the single logical qubit.
The primitive was proposed in \cite{brennen08}, to reconcile the strategy of MQC 
with the passive Hamiltonian protection of the resource ground state.
However, it has to be stressed our primitive does not fall into the {\it conventional} MQC where
quantum operations allowed for the computation are only the single-site measurements.
Since one of the strengths of the MQC lies in that every physical degree of freedom is
addressed only once and possibly destructively, i.e., disposable, here we assume this adiabatic 
weakening of the interaction before the measurement as a reasonable add-on requirement. 
The details of the scheme as well as how this primitive is conceivable in physical implementations
are found in \cite{brennen08} and references therein.
Multiple 1D chains are needed to be coupled in the similar way with the quantum circuit, to build 
a universal quantum computer. However, here by analyzing the 1D chain structure as its key constituent,
we like to highlight a potentially fundamental, computational mechanism, whose applicational range
turns out to be as ubiquitous as the region of the quantum phase.

We show explicitly that regardless of $\beta$, our primitive always induces the logical action 
exactly equivalent to the one at the AKLT point ($\beta = -\tfrac{1}{3}$), if the global $SU(2)$
symmetry, which is homomorphic to $SO(3)$, is maintained during the adiabatic turning off of 
the isotropic interaction $h_{j,j+1}$.
Since after the adiabatic turning off, ${\mathbf S}_{j}$ is not constrained by 
the Hamiltonian, the relevant ground subspace is spanned effectively by the spin $1$ and 
the newly emergent edge state of the spin $\tfrac{1}{2}$ in the remaining chain from the site $j+1$.
Based on the addition rule of the angular momenta by 
$\tfrac{1}{2} \otimes 1 = \tfrac{1}{2} \oplus \tfrac{3}{2}$,
there are two sectors of the total spin $S_{\rm tot}$ by $\tfrac{1}{2}$ and $\tfrac{3}{2}$.
However, the original logical information stored in the $S_{\rm tot} = \tfrac{1}{2}$ must be adiabatically transfered 
faithfully into the new sector of $S_{\rm tot} = \tfrac{1}{2}$ because of the conservation of $S_{\rm tot}$ and 
$S^z_{\rm tot}$ under the rotational invariance.
Using the Clebsch-Gordan decomposition,
the effective description, before the single-spin measurement at the site $j$, is
%\begin{widetext}
\begin{align}
\ket{\Psi(j)} \longmapsto & \sqrt{\frac{2}{3}} \left[ 
\ket{{\mathcal G}_{0} (j+1)} \otimes (\frac{a_0}{\sqrt{2}} \ket{0_{j}} + a_1 \ket{-1_{j}}) \right.
\nonumber\\ 
& \left. - \ket{{\mathcal G}_{1} (j+1)}\otimes (a_0 \ket{+1_{j}} + \frac{a_1}{\sqrt{2}} \ket{0_{j}}) \right],
\label{eq:unit}
\end{align}
%\end{widetext}
where the quantum number in the kets are the values of $S^z_j$.
The entanglement spectrum, which is the eigenvalues of the reduced density operator,
consists of $\tfrac{2}{3}$ and  $\tfrac{1}{3}$, 
{\it independently of the stored logical information $(a_0 , a_1)$}.
Thus the entanglement entropy is constant ($\approx$ 0.92 ebit), which guarantees the capability
of the logical action by the correlation to be measured in the second part.

Second, let us define an orthonormal local basis of our standard measurement of the spin $1$
(using the zero-eigenvalue states of $S^\alpha_j$ ($\alpha = x, y, z$)) as
$\bra{x} = \tfrac{-1}{\sqrt{2}}(\bra{1} - \bra{-1})$,
$\bra{y} = \tfrac{1}{\sqrt{2}}(\bra{1} + \bra{-1})$, and
$\bra{z} = \bra{0}$.
It is readily checked that the relative state of Eq.~(\ref{eq:unit}) for every outcome $\bra{x_j}, \bra{y_j}, \bra{z_j}$ is
\begin{equation}
\tfrac{1}{\sqrt{3}} X_L \ket{\Psi(j+1)},
\tfrac{1}{\sqrt{3}} X_L Z_L \ket{\Psi(j+1)},
\tfrac{1}{\sqrt{3}} Z_L \ket{\Psi(j+1)},
\end{equation}
respectively, where $X_L = |{\mathcal G}_{0}\rangle \langle{\mathcal G}_{1} | + 
|{\mathcal G}_{1}\rangle\langle{\mathcal G}_{0} |$
and $Z_L = |{\mathcal G}_{0}\rangle \langle {\mathcal G}_{0}| - |{\mathcal G}_{1}\rangle \langle {\mathcal G}_{1}|$.
Thus remarkably, the correlation we can get by our primitive is the same as what appears
in the MPS representation of the 1D AKLT ground state (see Eq.~(\ref{eq:AKLT})) and thus induces
the logical action same as at the AKLT point, although the exact description of the ground state 
is unknown here!
It is straightforward, following the AKLT scheme in \cite{brennen08}, to show how the arbitrary
time evolution of the single logical qubit can be simulated.
Since the arbitrary non-abelian $SU(2)$ operation can be decomposed by the rotations around two axes, 
such as $z$ and $x$, using the Euler decomposition, it is sufficient to confirm that
the single-site measurement in a rotated basis,
\begin{equation}
\{\bra{\gamma^{z}_{j} (\theta)} \}  = 
\{\tfrac{1}{2} [(1 \pm e^{i\theta}) \bra{x_j} + (1 \mp e^{i\theta}) \bra{y_j}], \bra{z_j}\},
\label{eq:zrotation}
\end{equation}
indeed provides the logical rotation around the $z$ axis by an arbitrary angle $\theta$, i.e., 
$R^{z}(\theta) = \kb{{\mathcal G}_0} + e^{i \theta} \kb{{\mathcal G}_1}$.
The rotation around the $x$ axis can be obtained by exchanging the roles of $\bra{x}$ and $\bra{z}$.

What makes the AKLT point ($\beta = -\tfrac{1}{3}$) special is the Hamiltonian 
is additionally {\it frustration free} (i.e. the global ground state also minimizes
the energy of every summand), so that the adiabatic turing off its summand does not cause 
any change of the ground state.
From this perspective, the previous AKLT scheme in \cite{brennen08} can be seen as 
the intersection between the conventional MQC on a static resource state and our current scheme 
that gradually modifies the correlation using the intrinsic Hamiltonian.
We must emphasize that the general ground state of Eq.~(\ref{eq:H}) 
does not simply work as a 1D quantum wire by the aforementioned measurements 
in the conventional framework of MQC, {\it i.e., if the whole Hamiltonian is off beforehand}. 
The role of the (generally frustrated) gapped Hamiltonian in the ground-code 
MQC is not simply to provide a stability to the bulk against local perturbations (e.g., \cite{yarotsky06}), 
but also to modify the correlation to the edge state suitably through adiabatically turning off.
On the other hand, inspired by the renormalization group, there is an advanced protocol \cite{noteMQC} 
in the conventional MQC, which enables us to use the ground state of the Haldane phase approximately 
as the logical wire on a longer distance scale.

Now let us turn toward symmetry protections to stabilize the Haldane phase 
\cite{gu09,pollmann09} and the computation on its edge.
In \cite{pollmann09}, it has been shown that the topological phase is protected as far as at least one of 
the following three discrete symmetries is present: (i) the time-reversal (TR) symmetry 
($S_k^{x,y,z} \mapsto - S_k^{x,y,z}$),
(ii) the ${\mathbb Z}_2 \times {\mathbb Z}_2$ symmetry by $\pi$ rotations around a pair of orthogonal axes
($S^{z}_k \mapsto S^{z}_k$, $S^{x,y}_k \mapsto - S^{x,y}_k $ for the case of the $z$ axis),
(iii) the inversion symmetry ($S^{x,y,z}_{k} \mapsto S^{x,y,z}_{N-k+1}$).
Note that the translational symmetry is irrelevant.
Although the edge states would not exist in a general Haldane phase protected only by the inversion 
symmetry \cite{gu09,pollmann09}, our computational protocol actively breaks the inversion symmetry, 
so that we cannot rely on it.
If we require either of the other two however, edge states are present, and that has justified our formulation 
here using them (In contrast, the so-called string order parameter is present only with the 
${\mathbb Z}_2 \times {\mathbb Z}_2$ symmetry).
We observe that the single-site measurement of Eq.~(\ref{eq:zrotation}) corresponds to the
projection to an eigenstate of the local Hermitian operator 
$m^z_j (\theta) = - [\cos \theta ((S^{x}_j)^2  - (S^{y}_j)^2 ) + \sin \theta (S^{x}_j S^{y}_j + S^{y}_j S^{x}_j) ]$,
which is invariant under TR or $\pi$ $z$-rotation symmetries.
Since the first part by turning off $h_{j,j+1}$ can respect them as well,
the whole process of the primitive is compatible with maintaining these discrete symmetries, so that 
it does not lift the topological stability of the Haldane phase.
For completeness to build a universal computer, a similar analysis can be made 
for the case of a logical two-qubit gate on multiple parallely-arrayed chains. A possible way is to let 
two spin $1$'s $A_j$ and $B_j$, each of which has been pulled out adiabatically, interact by 
$U_{A_j , B_j} = \openone - \tfrac{1}{2} (S^z + (S^z)^2 ) \otimes (S^z + (S^z)^2 )$ 
before the standard measurement \cite{brennen08}.
That corresponds to the projection in the second part by a two-local term
$U^{\dag} [m^z_{A_j} (0)\otimes m^z_{B_j}(0)] U$, which is still invariant under a combined symmetry
by TR and $\pi$ rotations of two ($x$ and $y$) axes, so that it is compatible, too.
Finally, every single step of our primitive is expected to be carried in a time independent of $N$.
This is because we could imagine the evolution by the primitive as essentially interpolating adiabatically  
turning off $h_{j,j+1}$ and turning on a perturbative term like $m^z_j (\theta)$, without breaking the key 
discrete symmetries.
Although the local gap on the spin $j$ is disappearing, that makes a new edge state emerging 
in the remaining chain from $j+1$, whose ground state is still gapped and two-fold degenerate.

{\it Open-chain scheme by an initial mixed state.---}
An issue of the open-chain scheme is that the initial resource
may not be a unique pure state under the situation of the preparation by the cooling,
because of the four-fold degeneracy of the ground state.
Although a global measurement of the total spin $S_{\rm tot}$ is sufficient to single out 
the singlet ground state, that may not be convenient in practice. 
Here we show how the proposed computation works without an initialization to a pure state.
According to our primitive, the logical action is always
exactly the same as if we were working on the AKLT ground state in the conventional MQC.
Thus, to see the effect of four-fold degeneracy explicitly, we may take advantage of 
the MPS form of the AKLT ground state \cite{AKLT87, arovas88, MPS92} (up to 
the normalization factor $\tfrac{1}{2 \sqrt{3^N}}$),
\begin{equation}
\ket{\Phi^{\mu \nu}} = \sum_{\alpha_k = x,y,z}  \!\!\!
{\rm tr}\left[(-Z)^\nu X^\mu \prod_{k = N}^{1} M[\alpha_k] \right] \ket{\alpha_1 \ldots \alpha_N} ,
\label{eq:AKLT}
\end{equation}
where $\mu, \nu = 0,1$ and $M[x] = X, M[y] = X Z, M[z] = Z$ in terms of the Pauli operators of size 
$2 \times 2$.  $\ket{\Phi^{0 0}}$ is the singlet ground state and the three others are the triplet. 
The convenience at the AKLT point comes from the feature that these Pauli operators 
$X$ and $Z$ can be {\it identified} with the aforementioned $X_L $ and $Z_L$ acting on the single 
edge state.

Suppose we are given a classical mixture of these four degenerate ground states with an arbitrary 
non-negative weight,
$\rho_{\mathcal G} = \sum_{\mu, \nu = 0,1} \lambda_{\mu \nu} \kb{\Phi^{\mu \nu}}$, as an initial
resource (it can be shown in general that any mixed state with
off-diagonal elements in $\rho_{\mathcal G}$ is fine as well).
To initialize or read out the logical wire, we could measure in the $S^z_j$ eigenbasis \cite{gross07}. 
If the outcome is successfully either $S^z_j = \pm 1$, a {\it known} fiducial state 
by the eigenstate of $Z$ is ``inserted'' (up to a so-called by-product operator which can be 
incorporated according to a standard adaption procedure of MQC (see, e.g., \cite{gross07,brennen08})),
otherwise we try the same at the next site.
With its expectation length finite, this protocol can create the region sandwiched by fixed ``boundaries'' 
in the product of the matrices $M$'s excluding $(-Z)^\nu X^\mu$.
The upshot is that, regardless of the ambiguity about $(\mu, \nu)$, 
we could use any finite connected region (say $k=\ell_{\rm s}, \ldots ,\ell_{\rm e}$ such that 
$|\ell_{\rm s}|, | N-\ell_{\rm e}| > \xi$) 
separated by these fixed boundaries 
for computation, and leave the initial mixedness to the rest of the unmeasured chain from $k= \ell_{\rm e} +1$. 
The protocol has also significance in the application of quantum error correction for fault tolerance, since 
it guarantees we do not have to measure the whole chain to read out the logical information.
It is interesting to mention that $\rho_{\mathcal G}$ is to some extent analogous to the mixture of 
the four Bell states (by two spin $\tfrac{1}{2}$'s). 
Although the latter would not be able to teleport the information perfectly anymore,
$\rho_{\mathcal G}$ is still capable of simulating a perfect logical wire.
This is because, based on the area law of entanglement, the bulk sustains much entanglement   
enough for $\rho_{\mathcal G}$ to be entangled even when a constant amount of entanglement 
near the boundaries was decohered initially.

{\it Discussion: holographic nature.---}
Now that we realize the whole analysis manifests a prevailing modern principle of physics, {\it holographic principle} 
\cite{holography}, in our model in that the logical information, which is capable of describing 
any dynamics of the ``universe'' by our computational primitive, is written on the edge, 
i.e., a holographic screen on the boundary.  
In this perspective, our boundary theory by the edge states is dual to the bulk theory by the MPS
representation of the global ground state in the previous work \cite{brennen08}, where the logical information
is {\it interpreted} to be encoded in the degenerate ground subspace.
However, the strength of the current boundary theory lies in that it provides a universal result
valid in the whole points of our isotropic Haldane phase, and suggests a broader computational potential 
even when the bulk theory cannot be exactly analyzed. Obviously, the ground state need not be described
by the valence bond solid \cite{AKLT87,arovas88, verstraete04}, that models the edge state explicitly.

One might wonder what makes the 1D Haldane phase special in our approach.
If the degeneracy of the ground state is desired to be robust without the symmetry protection, 
it seems to be necessary to consider the 2D (or 3D) topological orders.
Indeed, in a corresponding way to the vortices in the bulk, their edge states could
have Majorana fermionic zero modes, whose exotic statistics as anyons are famous
in the use for topological quantum computation \cite{kitaev03}.
Despite recent promising progress about the topological insulators \cite{hasan01} which are supposed 
to realize the Moore-Read state, 
the way to topological computation may not yet be plain because of the Ising anyons being 
short to universal computation, as well as of technical challenges to control anyons.
Here we intend to develop a computational mechanism, which is ubiquitous in a certain quantum 
phase and available in a relatively simple architecture.
In comparison to the topological computation where the bulk is merely a substrate to realize
the anyonic statistics, our model highlights the role of bulk entanglement,  
which persists through the adiabatic decoupling process from the boundary with 
the symmetry protection, and supports a computational capability on the edge state.
This is clearest at the AKLT point, as the ground state does not change and remains entangled 
between the spin $1$ pulled out and the bulk chain. Such a unique feature of the Haldane phase, 
in terms of the symmetry-protected entanglement spectrum, explains why our primitive does not work, 
for example, at the gapped dimer phase ($\beta > 1$) of Eq.~(\ref{eq:H}).

{\it Acknowledgment.---}
We acknowledge the valuable discussions 
with X.-G. Wen and H. Katsura about symmetry protections of topological orders, 
with D. Gross about the open-chain scheme, 
with S. Bartlett, G. Brennen, and J. Renes through the collaboration of \cite{noteMQC}.
The work is supported by the Government of Canada through Industry Canada and
by Ontario-MRI.


\begin{thebibliography}{99}

\bibitem{raussendorf01}
R. Raussendorf and H.J. Briegel, Phys. Rev. Lett. {\bf 86}, 5188 (2001).

\bibitem{brennen08}
G.K.~Brennen and A.~Miyake, Phys. Rev. Lett. {\bf 101}, 010502 (2008).

\bibitem{haldane83}
F.D.M. Haldane, Phys. Lett. A {\bf 93}, 464 (1983).

\bibitem{AKLT87}
I. Affleck {\it et al.}, Phys. Rev. Lett. {\bf 59}, 799 (1987).

\bibitem{MPS92}
M. Fannes, B. Nachtergaele, and R.F. Werner, Comm. Math. Phys. {\bf 144}, 443 (1992).

\bibitem{gu09}
Z.-C. Gu and X.-G. Wen, Phys. Rev. B {\bf 80}, 155131 (2009).

\bibitem{pollmann09}
F. Pollmann {\it et al.}, arXiv:0909.4059; 
Phys. Rev. B {\bf 81}, 064439 (2010).
%arXiv:0910.1811.

\bibitem{verstraete04}
F. Verstraete and J.I. Cirac, Phys. Rev. A {\bf 70}, 060302(R) (2004).

\bibitem{gross07}
D. Gross and J. Eisert, Phys. Rev. Lett. {\bf 98}, 220503 (2007);
D. Gross {\it et al.}, Phys. Rev. A {\bf 76}, 052315 (2007).

\bibitem{kennedy90}  
T. Kennedy, J. Phys. Cond. Mat. {\bf 2}, 5737 (1990).

\bibitem{hagiwara90}
M. Hagiwara {\it et al.}, Phys. Rev. Lett. {\bf 65}, 3181 (1990).

\bibitem{polizzi98}
E. Polizzi, F. Mila, and E.S. S{\o}rensen, Phys. Rev. B {\bf 58}, 2407 (1998).

\bibitem{yarotsky06}
D.A. Yarotsky, Comm. Math. Phys. {\bf 261}, 799 (2006).

\bibitem{noteMQC}
S.D. Bartlett, G.K. Brennen, A. Miyake, and J.M. Renes, arXiv:1004.4906.

\bibitem{arovas88}
D.P. Arovas, A. Auerbach, and F.D.M. Haldane, Phys. Rev. Lett. {\bf 60}, 531 (1988).

\bibitem{holography}
R. Bousso, Rev. Mod. Phys. {\bf 74}, 825 (2002);
J.D. Bekenstein, Sci. Amer. {\bf 289}, 58 (2003). 

\bibitem{kitaev03}
A.Y. Kitaev, Ann. Phys. {\bf 303}, 2 (2003). 

\bibitem{hasan01}
M.Z. Hasan and C.L. Kane, arXiv:1002.3895.

\end{thebibliography}
\end{document}